\documentclass[preprint, superscriptaddress, showpacs,preprintnumbers,amsmath,amssymb]{revtex4}
\usepackage{graphicx}

\begin{document}

\thispagestyle{empty}

\title{Demonstration of the Casimir force between
ferromagnetic surfaces of a Ni-coated sphere and
a Ni-coated plate}
\author{
A.~A.~Banishev,${}^1$
G.~L.~Klimchitskaya,${}^2$ V.~M.~Mostepanenko,${}^2$
and U.~Mohideen
}
\affiliation{\hspace*{-2mm}
Department of Physics and
Astronomy, University of California, Riverside, California 92521,
USA \\
${}^{\it 2}$Central Astronomical Observatory
at Pulkovo of the Russian Academy of Sciences,
St.Petersburg, 196140, Russia}

\begin{abstract}
We demonstrate the Casimir interaction between two ferromagnetic
boundary surfaces using the dynamic atomic force microscope.
The experimental data are found to be in excellent agreement with
the predictions of the Lifshitz theory for magnetic boundary
surfaces combined with the plasma model approach.
It is shown that for magnetic materials the role of
hypothetical patch potentials is opposite to that required for
reconciliation of the data with
the Drude model.
\end{abstract}

\maketitle

The Casimir effect \cite{1} is of much interest due to its
promising multidisciplinary applications in nanotechnology,
condensed matter physics, physics of elementary particles, and
in gravitation and cosmology \cite{2,3}.
Many experiments on measuring the Casimir force between
boundary surfaces made of different materials separated by a
vacuum gap or a liquid have been performed in the last
15 years \cite{4,5,6}. It was shown that the magnitude of
the Casimir force can be controlled by using different
boundary materials \cite{7,8}, phase transitions
\cite{9,10,11,12,13}, and by using the
boundary surfaces structured with nanoscale corrugations
\cite{14,15,16,17}.

A unified description of both
the van der Waals and Casimir forces is given by the
Lifshitz theory \cite{18} in terms of the
dielectric permittivity
$\varepsilon(\omega)$ and magnetic permeability
$\mu(\omega)$.
The role of magnetic materials in the Casimir force has
been studied theoretically
\cite{34,37,38,39,M1,M2,M3,M4,36,36a,M5,M6}.
The interest stems from the possibility to obtain a
repulsive Casimir force for application in micromachines.
Using real magnetic materials \cite{38,M3} did not validate the
early results which used constant $\varepsilon$ and $\mu$.
As $\mu(i\xi)$ can be large only at
$\xi<10^5-10^9\,$Hz,
its entire contribution to the Lifshitz formula is through
the zero Matsubara frequency \cite{36,36a}.
For metals, the zero-frequency term is strongly influenced
by the inclusion (Drude model approach) or neglect
(plasma model approach) of the relaxation properties of free
electrons \cite{4}. Thus using $\mu$ provides another
parameter to study the role of the relaxation properties of
free electrons in the Casimir effect.
Some experiments demonstrate
strong disagreement between the measured data and
theoretical predictions when the relaxation properties of
electrons are taken into account for metals
\cite{4,21,23} or the dc conductivity is
included for dielectrics \cite{4,12,13}.
The same data are found to be consistent with theory when
the relaxation properties are neglected for metals
 or the dc conductivity of dielectrics
is disregarded. Two other experiments \cite{28,29} are
claimed to be in favor of the Drude model approach
(see critical discussion in  \cite{30,31,32,33}).
It was also hypothesized \cite{patch}
that the effect of large patches might
bring the experimental data of Ref.~\cite{21} in agreement
with the predictions of the Drude model approach
(see also discussion in Ref.~\cite{23}).

In this Letter we describe demonstration of the Casimir force
between surfaces of a plate and a sphere, both  coated with
ferromagnetic metal Ni, performed by means of dynamic atomic
force microscope (AFM) using the frequency shift technique.
The Lifshitz theory was generalized for
the case of magnetic bodies in Ref.~\cite{34}, but
till now
was not unequivocally verified experimentally. Note that
measurements of the Casimir interaction between an Au-coated
sphere and a Ni-coated plate \cite{35} confirmed the influence
of magnetic properties on the Casimir force under an
assumption that the plasma model approach is adequate
(for Au interacting with Ni the Drude model approach is not sensitive to
magnetic properties and leads
to almost the same results as the
plasma model approach \cite{35}).
The advantage of Ni-Ni
test bodies used here is that the magnetic properties
significantly affect the Casimir force when both the
plasma and Drude model approaches
are used leading to considerably different
results \cite{36}. Using this property, we have unequivocally
confirmed that the magnetic properties influence the Casimir
force in accordance with predictions of the Lifshitz theory.
The agreement is excellent with
the plasma model approach, and the Drude model approach is
excluded by our data at a 95\% confidence level.
We have also excluded any possible role of patch effects on the
conclusions obtained. This opens opportunities for
far-ranging applications of the magnetic Casimir effect in
nanotechnology including the realization of the Casimir
repulsion through a vacuum gap \cite{M4,36,36a,M6}.

Here we have used the same apparatus and cantilever
preparation as in Refs.~\cite{23,35}.
The gradient of the Casimir force was measured acting
between a Ni-coated hollow glass microsphere of
$R=61.71\pm 0.09\,\mu$m radius attached to the tip of
a rectangular Si cantilever
and a Si plate also coated with Ni.
The thicknesses of Ni coating were $210\pm 1\,$nm and
$250\pm 1\,$nm on a sphere and a plate, respectively.
The hollow sphere leads to higher resonant frequencies and
mechanical $Q$-factors offering higher sensitivities.
To promote adhesion of the Ni coating a 10\,nm layer
of Cr followed by 40\,nm layer of Al was done first.
The coatings were performed at $10^{-6}\,$Torr.
To achieve uniformity of Ni layers, the sample was rotated during
evaporation of the metals. A coating rate
$\approx 3\,${\AA}/s was used.
Both test bodies were cleaned using a
multi-step procedure to remove any attached adsorbates
(both neutral and with net charge) and
debris (see \cite{23} for details).
The cantilever was clamped in a specially fabricated holder
and placed inside the vacuum chamber that was capable of reaching
a pressure of $10^{-9}\,$Torr by using mechanical, turbo and
ion pumps. The Ni-coated plate was fixed on the top of the
piezo with double sided vacuum adhesive tape. The movement
of the piezo was calibrated by a fiber interferometer with
$635.0\pm 0.3\,$nm laser source.

The dynamic measurement scheme in the
frequency modulation mode, as in Refs.~\cite{23,35,Ra}, was
used.
The directly measured quantity was the change of resonant
frequency of the periodically driven cantilever which was detected
by a phase locked loop system \cite{Rb,PLL} (see details
for our setup in Refs.~\cite{23,35}).
The driving frequency was kept near the resonance frequency
of the cantilever to obtain the highest signal to noise
ratio. The resonance frequency was detected with an optical
interferometer \cite{Rb,optInt}. To keep the interferometric cavity length
between the top of the cantilever and the end of the fiber fixed,
we used a piezo above the cantilever, which was controlled by
a proportional-integral-derivative feedback loop. This prevents
errors in the sphere-plate separation distance $a$ due to
cantilever deflection from Casimir, $F(a)$, and electrostatic,
$F_{\rm el}(a)$, forces.

For small oscillations in the presence of an external
force $F_{\rm tot}(a)=F_{\rm el}(a)+F(a)$, the
measured frequency shift
$\Delta\omega=\omega_r-\omega_0$ is expressed
\cite{23,35} as
$\Delta\omega=-(\omega_0/2k)F_{\rm tot}^{\prime}(a)$.
Here, $\omega_r$ is the resonance frequency in the presence
of $F_{\rm tot}$, $\omega_0$ is the natural resonance frequency,
$k$ is the spring constant of the cantilever, and
$a=z_{\rm piezo}+z_0$ ($z_{\rm piezo}$ is the plate
movement due to the piezoelectric actuator
which is calibrated interferometrically
 and $z_0$ is the
point of the closest approach between the two surfaces,
which in our case is much larger than the separation on
contact). The electric force
can be expressed as $F_{\rm el}(a)=X(a,R)(V_i-V_0)^2$,
where $X(a,R)$ is the known function \cite{3,4,23},
$V_i$ are the voltages applied to the plate, and $V_0$ is the
 residual potential difference.
In terms of the measured parameters, $\Delta\omega$
takes the form
\begin{equation}
\Delta\omega=-\beta(V_i-V_0)^2-
C F^{\prime}(a),
\label{eq1}
\end{equation}
\noindent
where $C=\omega_0/(2k)$ and
$\beta\equiv\beta(z_{\rm piezo},z_0,C,R)=CX^{\prime}(a,R)$.

A sufficiently precise electrostatic calibration, i.e.,
determination
of $V_0$, $z_0$, $C$, and $\beta$ from measurements of electric
forces is possible because we use a large perfectly shaped
sphere made from the liquid phase. The theoretical electric
force in the sphere-plane geometry is known exactly and the
potential between a sphere and a plane
can be precisely determined.
For electrostatic calibrations and measurements of $\Delta\omega$,
11 different voltages in the range from --64.5 to
31.6\,mV were applied to the Ni plate, while the sphere remained
grounded. The plate was moved toward the sphere starting at the
maximum separation of $2.3\,\mu$m and the corresponding
$\Delta\omega$
was recorded at every 0.14\,nm. Continuous triangular
voltages at 0.01\,Hz were applied to the tube piezo to move the
plate toward the sphere. This set of measurements was repeated
three times. The small mechanical drift 0.003\,nm/s in the
$z_{\rm piezo}$ was corrected as described in
Refs.~\cite{23,35}.
The parabolic dependence of $\Delta\omega$ on $V_i$ was used
to find $V_0$ at each separation \cite{13}.
Note that $V_0$ is separation independent indicating the
lack of any adverse surface contaminants and
the high quality of the measured
data (see Fig.~1 where the best fit of $V_0$ to the straight line
leads to a slope equal to only $1.5\times 10^{-5}\,$mV/nm).
The mean value of $V_0=-17.7\pm 1\,$mV was found.
The values of $z_0$ and $C$ were found by a least $\chi^2$
fitting of $\beta$ in Eq.~(\ref{eq1}).
The mean values are $z_0=221.1\pm 0.4\,$nm and
$C=52.4\pm 0.16\,$kHz\,m/N (the errors are indicated at a
67\% confidence level).
Then we obtain
$\beta=(\pi\epsilon_0CR/a^2)(1-2c_1a^2/R^2-4c_2a^3/R^3+\ldots)$,
where $c_1$ and $c_2$ are given in Ref.~\cite{3} and
$\epsilon_0$ is the permittivity of vacuum.
The absence of calibration errors in the obtained values of
 $z_0$ and $C$ was confirmed by their independence of the
separation  region used in calibration \cite{23,35}.
After the values of $z_0$ and $C$ were found, the measured
$\Delta\omega$ was converted into $F_{\rm tot}^{\prime}(a)$
and the absolute separation distances were
determined.

Now the 33 values of
$F^{\prime}(a)$ at each $a$ can be obtained from
Eq.~(\ref{eq1}) by subtracting the contribution of
$F_{\rm el}^{\prime}(a)$.
They are shown in Fig.~2 at $a$ from 223 to
320\,nm with a step of 2\,nm.
 The statistical properties of the
data are characterized by the histogram shown
in an inset to Fig.~2 at $a=250\,$nm. It is described by Gaussian
distribution with the standard deviation
$\sigma_{F^{\prime}}=0.92\,\mu$N/m and
mean
${F^{\prime}}=74.17\,\mu$N/m.
 The mean values of $F^{\prime}(a)$ as function of
$a$ (with a step of 1\,nm) are shown as crosses in Fig.~3(a--d)
where the arms of the crosses indicate the total experimental
errors found at a 67\% confidence level. The total errors are
mostly determined by the systematic error which are caused by the
errors in calibration. Thus, the systematic errors
in $F^{\prime}(a)$ at $a=223$, 250, 300, and 350\,nm are equal to
1.20, 1.05, 0.89, and $0.81\,\mu$N/m (i.e., 1.1\%, 1.4\%,
2.4\%, and 3.9\% of the force gradient), respectively.
These are quite sufficient to
discriminate between different theoretical predictions
(see below). The random error is equal to only $0.18\,\mu$N/m
and does not depend on $a$.

The experimental data for $F^{\prime}(a)$ were
compared with predictions of the Lifshitz theory.
The Lifshitz formula for magnetic materials
\cite{3,34,37,38,39} was adapted for sphere-plate geometry
using the proximity force approximation (this leads to
$<a/R$, i.e., $<0.36$\% error at the shortest
separation \cite{Fosco,40}) with the result
\begin{equation}
F^{\prime}(a)=2k_BTR\sum_{l=0}^{\infty}
{\vphantom{\sum}}^{\prime}\int_0^{\infty}q_l
k_{\bot}dk_{\bot}\sum_\alpha
\frac{r_{\alpha}^2}{e^{2aq_l}-r_{\alpha}^2}.
\label{eq2}
\end{equation}
\noindent
Here, $k_B$ is the Boltzmann constant, $T=300\,$K is
the temperature at the laboratory,
$q_l^2=k_{\bot}^2+\xi_l^2/c^2$, and
$\xi_l=2\pi k_BTl/\hbar$ with $l=0,\,1,\,2,\,\ldots$
are the Matsubara frequencies. The prime
multiplies the term with $l=0$ by
1/2 and the sum with respect to $\alpha$ implies
a summation in the transverse electric
($\alpha={\rm TE}$) and transverse magnetic
($\alpha={\rm TM}$) polarizations of the electromagnetic
field. The respective reflection coefficients are
given by
\begin{equation}
r_{\rm TM}=\frac{\varepsilon_lq_l-k_l}{\varepsilon_lq_l+k_l},
\quad
r_{\rm TE}=\frac{\mu_lq_l-k_l}{\mu_lq_l+k_l},
\label{eq3}
\end{equation}
\noindent
where $k_l^2=k_{\bot}^2+\varepsilon_l\mu_l\xi_l^2/c^2$,
and $\varepsilon_l\equiv\varepsilon(i\xi_l)$,
$\mu_l\equiv\mu(i\xi_l)$.

The permittivity $\varepsilon_l$ was obtained from the
optical data \cite{41} for the complex index of
refraction of Ni using the Kramers-Kronig relation. The data
were extrapolated to zero frequency either by means of the Drude
or the plasma
models. The plasma frequency
$\omega_p=4.89\,$eV
and the relaxation parameter
$\gamma=0.0436\,$eV
have been used \cite{41,42}. At $l=0$
the magnetic properties of Ni were described by the static
magnetic permeability $\mu_0=110$.
For all $l\geq 1$  at $T=300\,$K, $\mu_l=1$
because $\mu(\omega)$ rapidly falls to unity with
increasing $\omega$ \cite{36}.

The theoretical force gradients $F^{\prime}(a)$ were computed
using Eqs.~(\ref{eq2}) and (\ref{eq3}). The obtained values
were corrected for the presence of surface roughness.
The roughness profiles were investigated using an
AFM and the r.m.s.\ roughness on the sphere and the plate was
found to be $\delta_s=1.5\,$nm and $\delta_p=1.4\,$nm,
respectively. At separations $a\geq 223\,$nm this allows the use
of the multiplicative approach \cite{3,4,23,35}.
The theoretical results are shown in Figs.~2 and 3(a--d) within
different separation regions by the black and gray bands (their
widths are defined by the errors in the optical data) for the
Drude and plasma model approaches, respectively.
Note that at separations considered the difference between the
predictions of the Drude and plasma models for $F^{\prime}$ is
approximately proportional to $a^{-3}$.
As can be seen in Fig.~3, the Drude model approach is excluded by
the data at a 67\% confidence level over the region
from 223 to 420\,nm. The plasma model approach is in excellent
agreement with the data.
In Fig.~4 we plot the same data for $F^{\prime}$,
but with the total experimental errors determined at a 95\%
confidence level over the region from 223 to 350\,nm
(in the inset the interval from 300 to 350\,nm is shown on an
enlarged scale).
The errors at the 95\% confidence level are obtained in a
conservative way as the doubled errors found at the 67\%
confidence level \cite{Rabinovich}.
As can be seen in Fig.~4, over the region from
223 to 350\,nm the Drude model approach is excluded even at
a higher, 95\%, confidence level.
For $a$ from 420 to1000\,nm both the plasma and Drude model
approaches are consistent with the data. It should be noted,
however, that at large $a$ the data are not informative
with respect to the two models.
Thus, at $a=400$, 550, 750, and 1000\,nm the total relative
experimental error is equal to 6\%, 20\%, 100\%, and 321\%,
respectively, whereas the respective differences between the
predictions of the Drude and plasma model approaches are
equal to 8.5\%, 10\%, 12\%, and 14\%.

We emphasize that according to the Lifshitz theory the magnetic
properties of Ni in Ni-Ni system significantly influence the
gradient of the Casimir force in the framework of both
theoretical approaches (they increase $F^{\prime}$ when
the Drude model approach is used and decrease it
if the plasma model approach is applied).
Thus, our measurements unequivocally demonstrate the influence
of magnetic properties on the Casimir force as is predicted
by the Lifshitz theory combined with the plasma model approach.
Of even greater importance is the fact that for two magnetic
metals the Lifshitz theory predicts
$F_{\rm D}^{\prime}>F_{\rm p}^{\prime}$ where the
Drude and plasma model approaches are indicated
by the indices D and p (see Figs. 3 and 4 where the
black bands are above the gray). This is opposite to the
case of two nonmagnetic metals where
$F_{\rm D}^{\prime}<F_{\rm p}^{\prime}$ \cite{21,23}.
Thus, the inclusion of effect of patches
in the calculation  for two magnetic test
bodies is in principle incapable to bring the data in agreement
with the Drude model approach because patches always lead to an
additional attractive force. This proves that surface
patches do not play any role in our experiments and confirms
the model of patches \cite{43} which leads to a negligibly
small effect \cite{4}.

In this experiment both interacting bodies are magnetic
and consist of many domains.
Therefore it is necessary to analyze possible contribution of
magnetic forces into the measurement results. This is done by
considering two parallel Ni films of
$L_x\times L_y=0.9\times 1.1\,\mbox{cm}^2$ area and
applying the general formulation of the proximity force
approximation \cite{3,Derjaguin}.
For films more than 150\,nm thickness the
magnetization of each domain is perpendicular to the film
surfaces, i.e., has only the $z$-component equal to
$\pm M_s$, where $M_s=435\,\mbox{emu/cm}^3$ \cite{44,44a,44b}.
 The magnetization of
the first (1) and the second (2) films can be described
by a function of two variables $M_z^{(1,2)}(x,y)$.
In order to obtain the pair of infinite films described by the
periodic functions, we perform the periodic  continuation
of $M_z^{(1,2)}(x,y)$ as odd function with the periods
2$L_x$ and $2L_y$ and use the Fourier series
\begin{equation}
M_z^{(1,2)}(x,y)=\sum_{k=0}^{\infty}\sum_{n=0}^{\infty}
M_{kn}^{(1,2)}\sin\frac{k\pi x}{L_x}\sin\frac{n\pi y}{L_y}.
\label{eq4}
\end{equation}
\noindent
Here, $M_{00}^{(1,2)}\equiv 0$ if the spontaneous magnetization
is absent.

Next, using the standard formalism developed in magnetic force
microscopy \cite{45,46}, one can calculate the magnetic field
created by one Ni film and the magnetic force acting on
the other film. Keeping in mind that the magnetic force between
a pair of domains belonging to different films can be both
attractive and repulsive, and that these domains have different
size and are randomly arranged, the resulting magnetic force on
a sphere is equal to zero under the condition that the spontaneous
magnetization of at least one film is zero.
This conclusion is obtained for a film of infinitely large area.
Our Ni film of $L_x\times L_y$ area contains of about
$10^9$ domains whose sizes are approximately equal
to film thickness \cite{44} to minimize the magnetic energy.
In this case a noncompensated gradient
of the magnetic force is estimated to be less than
$10^{-2}\,\mu$N/m, i.e., a factor of 100 less than the
experimental error.

To avoid the spontaneous magnetization of Ni films, we made them
sufficiently thick and screened the weak environmental magnetic field
in our setup. If, however, there is some nonzero
spontaneous magnetization of both films, the resulting gradient of
the magnetic force acting on a sphere, although nonzero, is
negligibly small. This is because the magnetic field near the
center of a large film does not depend on $z$ for $z\ll L_x,L_y$
\cite{47}. For example, even for a fully magnetized film
(which is not the case for our setup) the
gradient of the magnetic force acting on a sphere in the region
of experimental separations is much less than
$2\times 10^{-3}\,\mu$N/m, i.e., much less than the experimental
error in the measurements of $F^{\prime}(a)$.

To conclude, we have experimentally demonstrated that the magnetic
properties of Ni influence the Casimir
interaction as predicted by the Lifshitz theory combined with the
plasma model approach. The Drude model approach in application to
magnetic metals is excluded at a 95\% confidence level.
We have also shown that any hypothetical patch potential
will only exacerbate the deviation from the Drude model approach.
The obtained results allow realization of
the Casimir repulsion through a vacuum gap which could lead
to many potential applications in nanotechno\-lo\-gy.

%%%%%%%%%%%%%%%%%%%%%%%%%%%%%%%%%%%%
\section*{Acknowledgments}
This work was supported by the DOE Grant
No.~DEF010204ER46131 (equipment, G.L.K., V.M.M., U.M.)
and NSF Grant
No.~PHY0970161 (G.L.K., V.M.M., U.M.).
G.L.K.\ and V.M.M.\ were also partially supported
by the DFG grant BO\ 1112/21--1.
%%%%%%%%%%%%%%%%%%%%%%%%%%%%%%%%

%%%%%%%%%%%%%%%%%%%%%%%%%%%%%%%%%%%%%%%%
%\end{document}
%%%%%%%%%%
%%%%%%%__FIGURE__1__%%%%%%%%%%%%%%%%%%%%
\begin{figure}[t]
\vspace*{-7cm}
\centerline{\hspace*{3cm}
\includegraphics{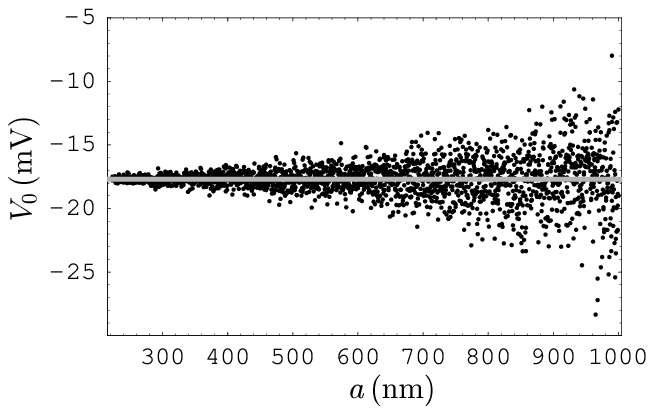}
}
\vspace*{-16cm}
\caption{The residual potential difference between a
Ni-coated sphere and a Ni-coated plate as a function of
separation. The mean value of $V_0$ is shown by the gray line.
}
\end{figure}
%%%%%%%%%%%%%%
%%%%%%%__FIGURE__2__%%%%%%%%%%%%%%%%%%%%
\begin{figure}[b]
\vspace*{-7cm}
\centerline{\hspace*{3cm}
\includegraphics{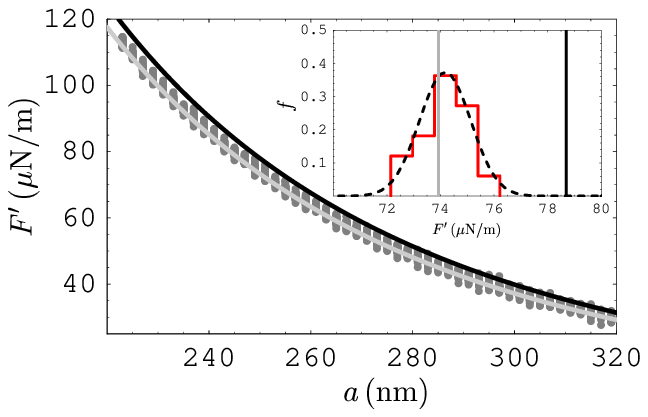}
}
\vspace*{-15cm}
\caption{(Color online)
Comparison between the nonaveraged experimental data for
$F^{\prime}$ (gray dots) and theory
(black and gray bands computed using the Drude
and plasma model approaches, respectively).
The inset shows the
histogram for the measured $F^{\prime}$ at
$a=250\,$nm. $f$ is the fraction of 33 data points
having the values of $F^{\prime}$ in the bin indicated
by the respective vertical lines. The corresponding
Gaussian distribution is shown by the dashed line. The
black and gray vertical lines show the theoretical
predictions of the Drude and plasma model approaches.
}
\end{figure}
%%%%%%%%%%%%%%
%%%%%%%__FIGURE__3__%%%%%%%%%%%%%%%%%%%%
\begin{figure}[t]
\vspace*{-1cm}
\centerline{\hspace*{1cm}
\includegraphics{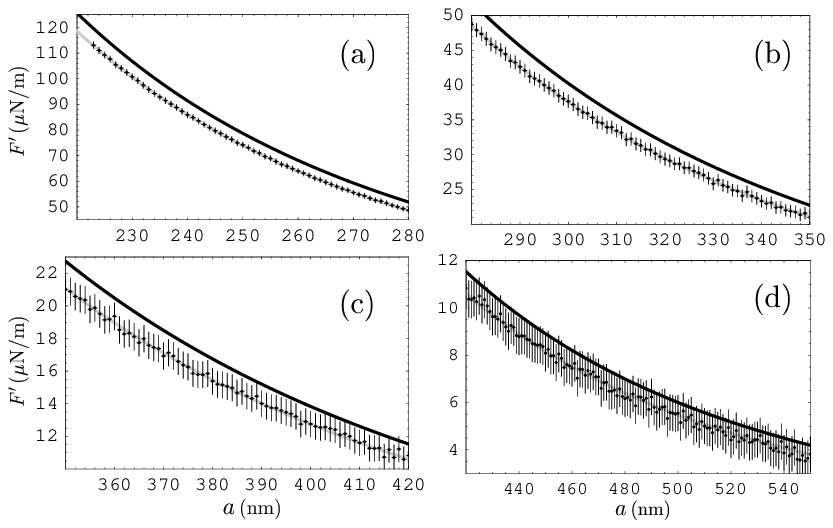}
}
\vspace*{-19cm}
\caption{Comparison between the experimental data for
$F^{\prime}$ (crosses plotted at a 67\% confidence level)
and theory (black and gray bands computed using the Drude
and plasma model approaches, respectively).
}
\end{figure}
%%%%%%%%%%%%%%

%%%%%%%__FIGURE__4__%%%%%%%%%%%%%%%%%%%%
\begin{figure}[b]
\vspace*{-6cm}
\centerline{\hspace*{3cm}
\includegraphics{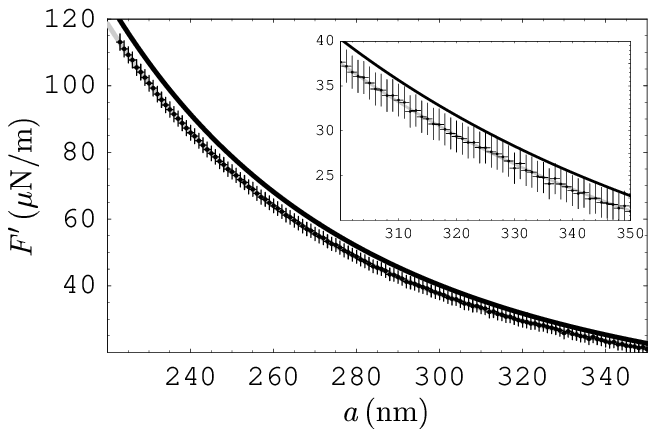}
}
\vspace*{-16cm}
\caption{Comparison between the experimental data for
$F^{\prime}$  plotted at a 95\% confidence level
and theory (black and gray bands computed using the Drude
and plasma model approaches, respectively).
}
\end{figure}
%%%%%%%%%%%%%%
%%%%%%%%%%%%%%%%

\begin{thebibliography}{99}
\bibitem {1}
H.~B.~G.~Casimir,
{ Proc. K. Ned. Akad. Wet. B}
{\bf 51}, 793 (1948).
\bibitem{2}
K.~A.~Milton,
{\it The Casimir Effect: Physical Manifestations of
Zero-Point Energy}
(World Scientific, Singapore, 2001).
\bibitem{3}
M.~Bordag, G.~L.~Klimchitskaya, U.\ Mohideen, and
V.\ M.\ Mostepanenko, {\it Advances in the Casimir Effect}
(Oxford University Press, Oxford, 2009).
\bibitem{4}
G.~L.~Klimchitskaya, U. Mohideen, and V.\ M.\ Mostepanenko,
 Rev. Mod. Phys. {\bf 81}, 1827 (2009).
\bibitem{5}
G.~L.~Klimchitskaya, U. Mohideen, and V.\ M.\ Mostepanenko,
 Int. J. Mod. Phys. B {\bf 25}, 171 (2011).
\bibitem{6}
A.~W.~Rodriguez, F.~Capasso, and S.~G.~Johnson,
Nature Photon. {\bf 5}, 211 (2011).
\bibitem{7}
F.~Chen, G.~L.~Klimchitskaya,
V.\ M.\ Mos\-te\-pa\-nen\-ko and U.~Mohideen,
{Phys. Rev. Lett.} {\bf 97}, 170402 (2006).
\bibitem{8}
S.~de~Man, K.~Heeck, R.~J.~Wijngaarden, and D.~Iannuzzi,
Phys. Rev. Lett. {\bf 103}, 040402 (2009).
\bibitem{9}
G.~Torricelli, P.~J.~van~Zwol, O.~Shpak, C.~Binns, G.\ Palasantzas,
B.~J.~Kooi, V.\ B.\ Svetovoy and M.\  Wuttig,
{Phys. Rev.} A {\bf 82}, 010101(R) (2010).
\bibitem{10}
G.~Bimonte,
Phys. Rev. A {\bf 78}, 062101 (2008).
\bibitem{11}
G.~Bimonte, E.~Calloni, G.\ Esposito, L.\ Milano, and
L.\ Rosa,
Phys. Rev. Lett. {\bf 94}, 180402 (2005).
\bibitem{12}
C.-C.~Chang, A.~A.~Banishev,
G.~L.~Klimchitskaya, V.\ M.\ Mostepanenko, and U.\ Mohideen,
Phys. Rev. Lett.  {\bf 107}, 090403 (2011).
\bibitem{13}
A.~A.~Banishev, C.-C.~Chang, R.~Castillo-Garza,
G.~L.~Klimchitskaya, V.\ M.\ Mostepanenko, and U.\ Mohideen,
Phys. Rev. B {\bf 85}, 045436 (2012).
\bibitem{14}
H.~B.~Chan, Y.~Bao, J.\ Zou, R.\ A.\ Cirelli, F.\ Klemens,
W.\ M.\ Mansfield and C.\ S.\ Pai,
{ Phys. Rev. Lett.} {\bf 101}, 030401 (2008).
\bibitem{15}
Y.~Bao, R.~Gu\'{e}rout, J.~Lussange, A.\ Lambrecht, R.\ A.\ Cirelli, F.\ Klemens,
W.\ M.\ Mansfield, C.\ S.\ Pai
and H.~B.~Chan,
{Phys. Rev. Lett.} {\bf 105}, 250402 (2010).
\bibitem{16}
H.-C.\ Chiu,  G.~L.~Klimchitskaya, V.\ N.\ Marachevsky,
V.\ M.\ Mos\-te\-pa\-nen\-ko, and U.~Mohideen,
Phys. Rev. B {\bf 80}, 121402(R) (2009).
\bibitem{17}
H.-C.\ Chiu,  G.~L.~Klimchitskaya, V.\ N.\ Marachevsky,
V.\ M.\ Mos\-te\-pa\-nen\-ko, and U.~Mohideen,
Phys. Rev. B {\bf 81}, 115417 (2010).
\bibitem{18}
E.~M.~Lifshitz,
Zh. Eksp. Teor. Fiz. {\bf 29}, 94 (1955)
[Sov. Phys. JETP  {\bf 2}, 73 (1956)].
\bibitem{34}
P.~Richmond and B.~W.~Ninham,
J. Phys. C: Solid St. Phys. {\bf 4}, 1988 (1971).
\bibitem{37}
S.~Y.~Buhmann,  D.-G.~Welsch, and T.\ Kampf,
{Phys. Rev.} A {\bf 72}, 032112 (2005).
\bibitem{38}
M.~S.~Toma\v{s}, Phys. Lett. A {\bf 342}, 381 (2005).
\bibitem{39}
S.~J.~Rahi, T.~Emig, N.~Graham, R.\ L.\ Jaffe, and M.\ Kardar,
Phys. Rev. D {\bf 80}, 085021 (2009).
\bibitem{M1}
Yu.~S.~Barash and V.~L.~Ginzburg,
Usp. Fiz. Nauk {\bf 116}, 5 (1975)
[Sov. Phys. Usp. {\bf 18}, 305 (1975)].
\bibitem{M2}
O.~Kenneth, I.~Klich, A.\ Mann, and M.\ Revzen,
Phys. Rev. Lett. {\bf 89}, 033001 (2002).
\bibitem{M3}
D.~Iannuzzi and F.~Capasso,
Phys. Rev. Lett. {\bf 91}, 029101 (2003).
\bibitem{M4}
F.~S.~S.~Rosa, D.~A.~R.~Dalvit, and P.~W.~Milonni,
Phys. Rev. A  {\bf 78}, 032117 (2008).
\bibitem{36}
B.~Geyer, G.~L.~Klimchitskaya, and V.~M.~Mostepanenko,
Phys. Rev. B  {\bf 81}, 104101 (2010).
\bibitem{36a}
G.~L.~Klimchitskaya, B.~Geyer, and V.~M.~Mostepanenko,
 Int. J. Mod. Phys. A  {\bf 25}, 2293 (2010).
\bibitem{M5}
N.~Inui,
Phys. Rev. A  {\bf 84}, 052505 (2011).
\bibitem{M6}
N.~Inui,
Phys. Rev. A  {\bf 86}, 022520 (2012).
\bibitem{21}
R.~S.~Decca, D.~L\'opez, E.~Fischbach, G.~L.~Klimchitskaya,
 D.~E.~Krause, and V.~M.~Mostepanenko,
Phys. Rev. D {\bf 75}, 077101 (2007).
\bibitem{23}
C.-C.~Chang, A.~A.~Banishev, R.~Castillo-Garza,
G.~L.~Klimchitskaya, V.\ M.\ Mostepanenko, and U.\ Mohideen,
Phys. Rev. B {\bf 85}, 165443 (2012).
\bibitem{28}
 A.~O.~Sushkov, W.~J.~Kim, D.\ A.\ R.\ Dalvit,
and S.\ K.\ Lamoreaux,
Nature Phys. {\bf 7}, 230 (2011).
\bibitem{29}
D.~Garcia-Sanches, K.~Y.~Fong, H.~Bhaskaran, S.~Lamoreaux, and
H.~X.~Tang,
Phys. Rev. Lett. {\bf 109}, 027202 (2012).
\bibitem{30}
V.~B.~Bezerra, G.~L.~Klimchitskaya, U.\ Mohideen, V.~M.~Mostepanenko,
and C.~Romero,
{Phys. Rev. B} {\bf 83}, 075417 (2011).
\bibitem{31}
G.~L.~Klimchitskaya, M.~Bordag,  E.~Fischbach,
 D.~E.~Krause, and V.~M.~Mostepanenko,
 Int. J. Mod. Phys. A {\bf  26}, 3918 (2011).
\bibitem{32}
G.~L.~Klimchitskaya, M.~Bordag, and V.~M.~Mostepanenko,
 Int. J. Mod. Phys. A {\bf  27}, 1260012 (2012).
\bibitem{33}
M.~Bordag, G.~L.~Klimchitskaya, and V.~M.~Mostepanenko,
Phys. Rev. Lett. {\bf 109}, 199701 (2012).
\bibitem{patch}
R.~O.~Behunin, F.~Intravaia, D.~A.~R.~Dalvit, P.~A.~Maia Neto,
and S.~Reynaud, Phys. Rev. A {\bf 85}, 012504 (2012).
\bibitem{35}
A.~A.~Banishev, C.-C.~Chang,
G.~L.~Klimchitskaya, V.\ M.\ Mostepanenko, and U.\ Mohideen,
Phys. Rev. B {\bf 85}, 195422 (2012).
\bibitem{Ra}
T.~R.~Albrecht, P.~Gr\"{u}tter, D.~Horne, and D.~Rugar,
J. Appl. Phys. {\bf 69}, 668 (1991).
%\bibitem{Du}
%U.~D\"{u}rig, O.~Z\"{u}ger, and A.~Stalder,
%J. Appl. Phys. {\bf 72}, 1778 (1992).
\bibitem{PLL}
F.~J.~Giessibl,
Rev. Mod. Phys. {\bf 75}, 949 (2003).
\bibitem{Rb}
D.\ Rugar, H.~J.~Mamin, and P.~Guethner,
Appl. Phys. Lett. {\bf 55}, 2588 (1989).
\bibitem{optInt}
B.~C.~Stipe, H.~J.~Mamin, T.~D.~Stowe, T.\ W.\ Kenny,
and D.\ Rugar,
Phys. Rev. Lett. {\bf 87}, 096801 (2001).
\bibitem{Fosco}
C.~D.~Fosco, F.~C.~Lombardo, and F.~D.~Mazzitelli,
Phys. Rev. D {\bf 84}, 105031 (2011).
\bibitem{40}
G.~Bimonte, T.~Emig, and M.~Kardar,
Appl. Phys. Lett. {\bf 100}, 074110 (2012).
\bibitem {41}
{\it Handbook of Optical Constants of Solids},
ed. E.~D.~Palik (Academic, New York, 1985).
\bibitem{42}
M.~A.~Ordal, R.~J.~Bell, R.~W.~Alexander Jr., L.\ L.\ Long,
and M.\ R.\ Querry,
Appl. Opt. {\bf 24}, 4493 (1985).
\bibitem{Rabinovich}
S.~G.~Rabinovich,
{\it Measurement Errors and Uncertainties.
Theory and Practice}
(Springer-Verlag, New York, 2000).
\bibitem{43}
C.~C.~Speake and C.~Trenkel,
Phys. Rev. Lett. {\bf 90}, 160403 (2003).
\bibitem{Derjaguin}
B.~V.~Derjaguin, Kolloid. Z. {\bf 69}, 155 (1934).
\bibitem{44}
S.~Hameed,  P.~Talagala, R.~Naik, L.\ E.\ Wenger, V.\ M.\ Naik,
and  R.\ Proksch,
Phys. Rev. B {\bf 64}, 184406 (2001).
\bibitem{44a}
W.~H.~Kraan and M.~Th.~Rekveldt,
J. Magnetism \& Magn. Mat. {\bf 5}, 247 (1977).
\bibitem{44b}
O.~V.~Snigirev, K.~E.~Andreev, A.~M.~Tishin,
S.\ A.\ Gudoshnikov, and J.\ Bohr,
Phys. Rev. B {\bf 55}, 14429 (1997).
\bibitem{45}
A.~Wadas and  P.~Gr\"{u}tter,
Phys. Rev. B  {\bf 39}, 12014 (1989).
\bibitem{46}
D.~Sarid,
{\it Scanning Force Microscopy. With Application to Electric,
Magnetic and Atomic Forces}
(Oxford University Press, Oxford, 1994).
\bibitem{47}
J.~J.~S\'{a}enz, N.~Garcia, P.~Gr\"{u}tter, E.\ Meyer,
H.\ Heinzelmann,  R.\ Wiesendanger, L.\ Rosenthaler,
H.\ R.\ Hidber, and H.-J.\ G\"{u}ntherodt,
J. Appl. Phys. {\bf  62}, 4293 (1987).
\end{thebibliography}
\end{document}